\documentclass[prl, twocolumn, floatfix]{revtex4}
\setcitestyle{super}

\usepackage{graphicx}
\usepackage{hyperref}
\usepackage{amsmath}
\usepackage{color}
\usepackage[normalem]{ulem}

\begin{document}

\title{Kibble-Zurek mechanism in the self-organization of a cold atomic cloud}

\author{G. Labeyrie\footnote{To whom correspondence should be addressed.} and R. Kaiser}
\affiliation{Universit\'{e} C\^{o}te d'Azur, CNRS, INLN, 06560 Valbonne, France}

\begin{abstract}
When applying two counter-propagating laser beams to a cold cloud of Rubidium atoms, we observe the spontaneous formation of intensity patterns associated with a spatial structuration of the atomic spins. We study the average number of defects in these patterns as a function of the sweep time employed to cross the transition threshold. We observe a power-law decrease of the number of defects with increasing sweep time, typical of the Kibble-Zurek mechanism. The measured exponent is consistent with the prediction for a supercritical bifurcation.

\end{abstract}

\maketitle

Out of equilibrium dynamics is important in many aspects and has consequences which often cannot be neglected even for long time behavior. For instance the formation of topological defects after a quench across a phase transition is dependent on the out of equilibrium dynamics during such a transition. One elegant approach to describe the formation of such defects using parameters from equilibrium dynamics is the Kibble-Zurek mechanism (KZM), first introduced in cosmology in the context of the expansion of the early universe with fluctuations still visible today~\cite{Kibble1976, Zurek1985}. An important result of the KZM theory is a prediction for second-order phase transitions of the scaling of the density of defects with the speed of the quench~\cite{Zurek1985}. The KZM scheme has then been shown to be applicable to a larger variety of systems and thus accessible in laboratory experiments, making this a testable and useful approach to control the dynamics of a large class of transitions, including classical and quantum phase transitions \cite{Zurek2002}, cross overs and bifurcations \cite{Zurek2000, Casado2007, Miranda2013}. KZM has been investigated in a variety of experimental systems including liquid crystals~\cite{Yurke1991, Arecchi1999}, $^3$He\cite{Pickett1999}, superconductors\cite{Monaco2009}, cold ions~\cite{Singer2013}, and ultracold atoms~\cite{Sadler2006, Weiler2008, Chen2011, Ferrari2013, Braun2014, Corman2014, Navon2015, Dalibard2015}.

\begin{figure}
\begin{center}
\includegraphics[width=1.0\columnwidth]{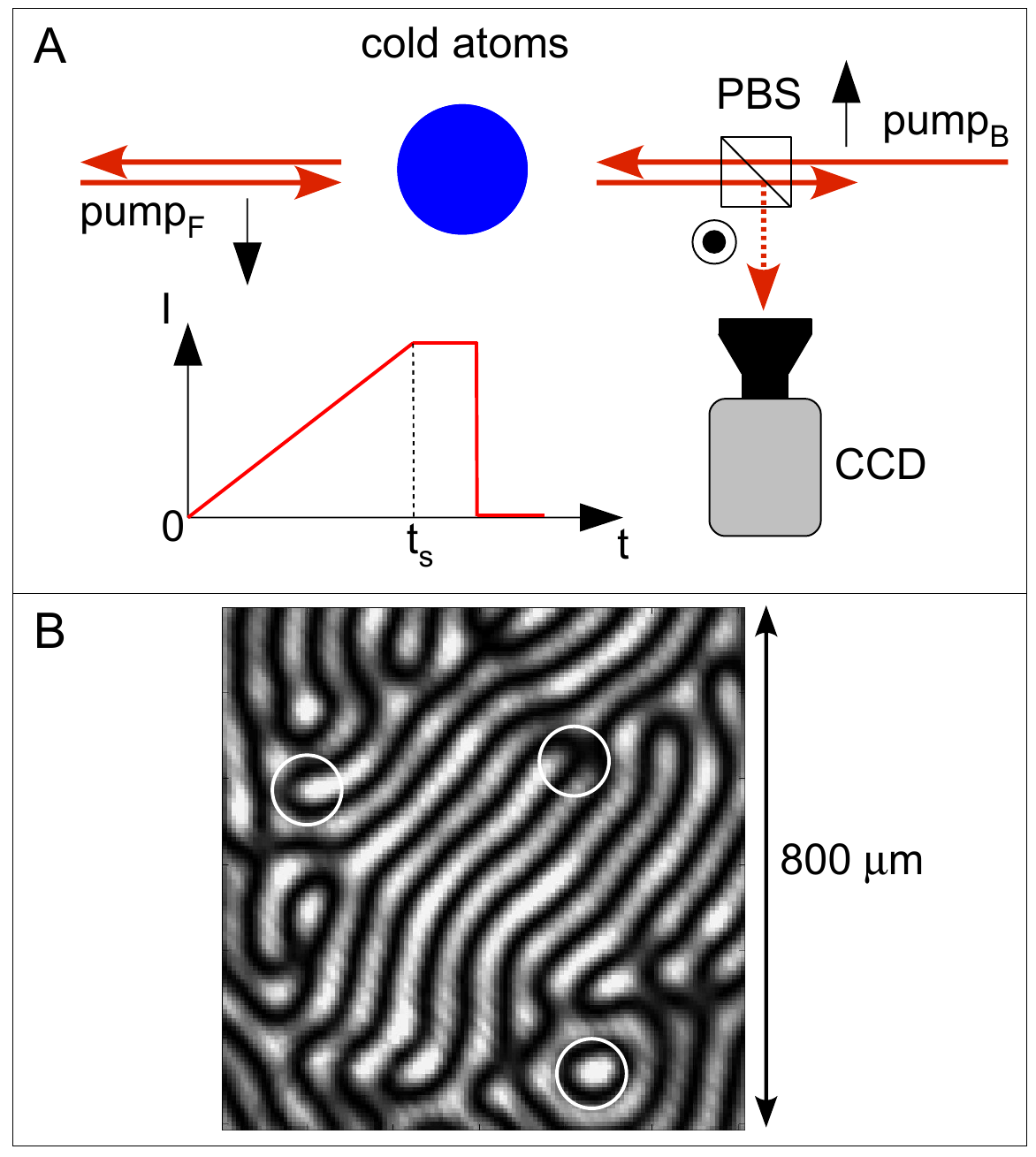}
\caption{(Color on line) Experimental configuration. (A) Two linearly-polarized, counter-propagating laser beams are applied to a cloud of cold atoms. The CCD images the transverse intensity distribution of the light in the polarization channel orthogonal to that of the pump (see text). The intensity of the pump is swept across the transition in a time $t_s$. (B) Typical image recorded on the CCD. The circles outline different kinds of defects.}
\label{fig1}
\end{center}
\end{figure}

For the past four years, we and co-workers have been investigating pattern-forming instabilities in cold atomic gases~\cite{Labeyrie2014, Labeyrie2015, Firth2016}. These experiments follow precursory results obtained in the nineties by several groups using hot atomic vapors as a nonlinear medium\cite{Grynberg1994, Ackemann1994, Ackemann1995}. The general principle is to apply to the atomic medium a retro-reflected laser beam (hereafter referred to as ''the pump'') detuned from the atomic resonance, in the so-called single mirror feedback configuration introduced by Firth in 1990~\cite{Firth1990}. As a result of the nonlinear interaction between light and matter, high-contrast spatial patterns develop both in the susceptibility of the atomic medium as well as in the electromagnetic field in the transverse plane, orthogonal to the propagation axis of the beams. Because the medium is invariant by translation and rotation about this axis, two continuous symmetries are spontaneously broken in the pattern formation process. All these experiments are performed by abruptly turning on the control field, here the laser beam, and the situation is thus that of a quench across a bifurcation. In most situations, we observe shot to shot fluctuations in the position and orientation of the pattern (this supposes a clean beam profile to avoid pinning), as well as the presence of domains and defects. The aim of this Letter is to investigate the role of the KZM scenario in the formation of these defects. To achieve this, we will study in the following the evolution of the average number of defects as we vary the sweep time of the control field across the transition.

A very specific feature of cold atomic clouds is that several distinct mechanisms are available to obtain the optical nonlinearity required for pattern formation~\cite{Labeyrie2014, Labeyrie2015, Firth2016}. Furthermore, as we discuss below, these mechanisms can be selected independently in the same experimental setup. In the search for KZM presented in this paper, we thus selected a specific nonlinearity which allowed us both to perform a study of the number of defects, and whose time constant was short enough to have access to a sufficient range of sweep times across the transition.

We shortly recall here the mechanisms that may lead to pattern formation in cold atoms. The first and conceptually simplest mechanism is the Kerr effect due to the saturation of a two-level atomic transition~\cite{Labeyrie2015}. Due to its simplicity, this nonlinear mechanism was favored in theoretical treatments of such instabilities~\cite{Grynberg1988a, Firth2016}. The corresponding ''2-level instability'' has a relatively high intensity threshold, since the saturation parameter
\begin{equation}
s = \frac{I}{I_{sat}} \frac{1}{1+4 (\delta / \Gamma)^2}
\label{s}
\end{equation}
describing the magnitude of the atom-field coupling needs to be of the order of unity. In this expression, $I$ is the laser intensity, $I_{sat} = 3.59$ mW/cm$^2$ the saturation intensity (taking into account the Zeeman structure), $\delta$ the frequency detuning between the laser and the atomic transition and $\Gamma = 2\pi \times 6.06$ MHz the atomic linewidth. When $s > 1$, the time constant for the nonlinearity approaches the excited state lifetime $\tau_{nat} = 27$ ns. Our previous observations suggest that pattern formation requires the nonlinearity to be self-focusing~\cite{Labeyrie2015}, i.e. that the refractive index increases with laser intensity. As a consequence, the 2-level instability is observed on the blue side of the transition ($\delta > 0$) only.

\begin{figure}
\begin{center}
\includegraphics[width=1\columnwidth]{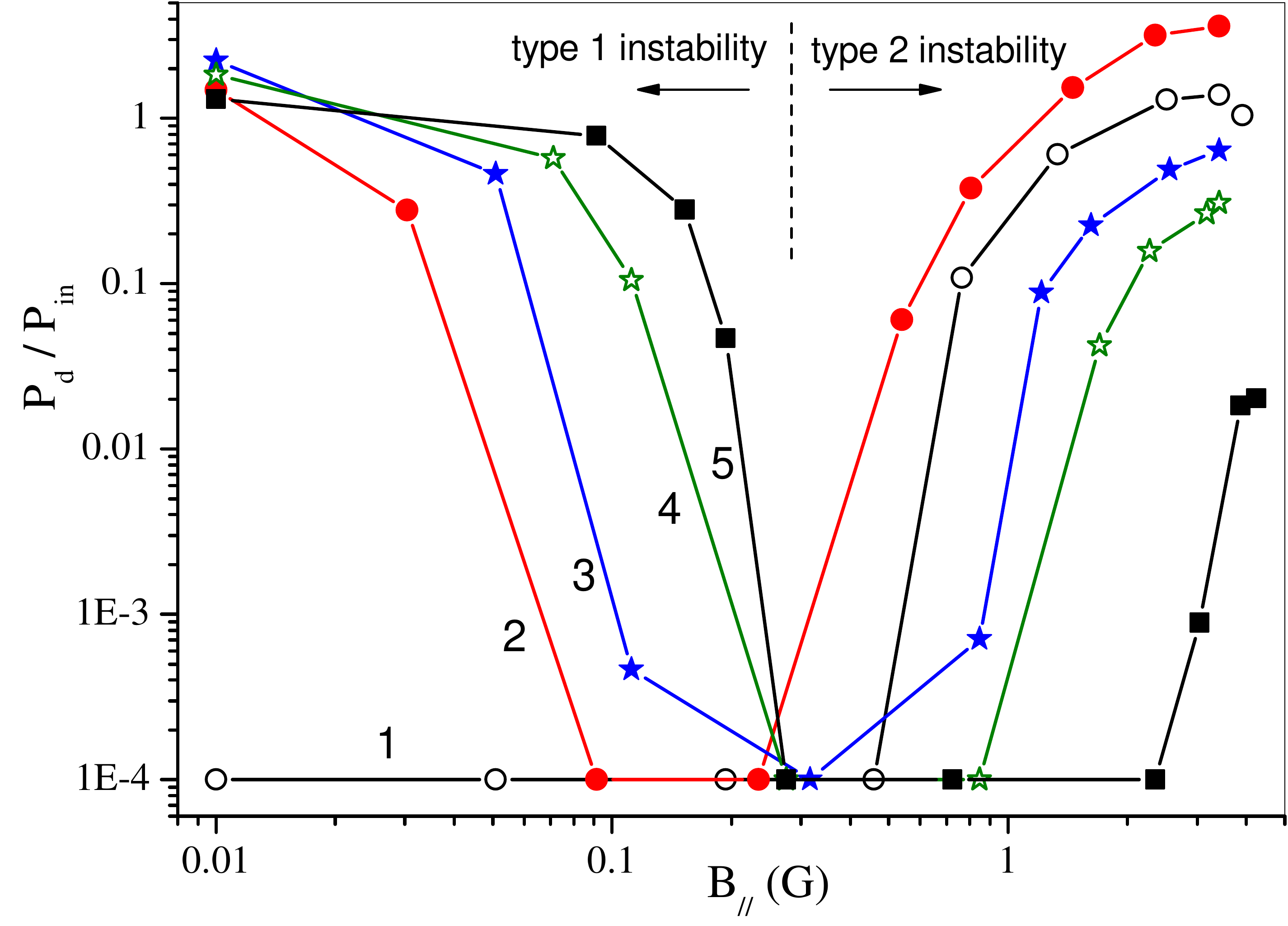}
\caption{Magnetic field dependence of the spin instability. We plot the diffracted intensity detected in the orthogonal channel (see text), as a function of the magnetic field applied along the polarization direction of the pumps. The various curves correspond to different values of the saturation parameter: (1) $s = 2.5\times10^{-3}$; (2) $s = 3.2\times10^{-3}$; (3) $s = 6\times10^{-3}$; (4) $s = 9\times10^{-3}$; (5) $s = 1.3\times10^{-2}$.}
\label{fig2}
\end{center}
\end{figure} 

A second mechanism, specific to cold atoms, is the spatial bunching of the atoms under the action of the dipole force. This leads to the ''optomechanical instability'' described in~\cite{Labeyrie2014}. This mechanism is slower (typically several tens of $\mu$s in our experimental conditions) since it requires the atoms to move over distances of the order of several tens or hundreds of $\mu$m. Because of the efficiency of spatial bunching, the intensity threshold of the optomechanical instability is lower than that of the 2-level instability. Since the efficiency of spatial bunching depends on the temperature of the atomic ensemble, this threshold also depends on temperature. The optomechanical instability is observed essentially for $\delta > 0$. The two instabilities described up to now are essentially scalar (no polarization dependence) and independent of the magnetic field. In both cases, the symmetry of the patterns is hexagonal.

The third mechanism, which is exploited in this work, is based on optical pumping between Zeeman sub-states and thus relies on spin degrees of freedom. It is this kind of nonlinearity that allowed the observation of patterns in hot atomic vapors in early experiments~\cite{Grynberg1994, Ackemann1994, Ackemann1995}. The appearance of high-contrast patterns in the transverse intensity profile of the beams is accompanied by the establishment of a strong spatial modulation of the Zeeman populations (and/or coherences), i.e. a \emph{spontaneous magnetic ordering} in the cold atomic sample~\cite{Firth2016, Kresic2016}. The main difference between this ''spin instability'' and the previous ones is its vectorial character, i.e. it depends on the polarization of the pump and the light polarization may not be preserved during the nonlinear interaction. Thus, the spatial instability is often accompanied by a \emph{polarization instability}. Another characteristic feature of the spin instability is its strong sensitivity to magnetic fields, which will be discussed below. Since Zeeman pumping only requires the absorption of a few photons, the intensity threshold of the spin instability can be very low: we have observed pattern formation for a saturation parameter well below $s = 10^{-3}$. This spin instability can be fast (time scale of the order of $\tau_{nat}$ if $s > 1$), and leads to the formation of patterns with various symmetries determined by the magnetic field and the laser intensity~\cite{Kresic2016}. Furthermore, Zeeman pumping towards stretched states leads to a self-focusing nonlinearity for $\delta < 0$. As a result, we only observe the spin instability for a red detuning ($\delta < 0$) which readily allows to separate it from the two previous ones.

\begin{figure}
\begin{center}
\resizebox{1.0\columnwidth}{!}{\includegraphics{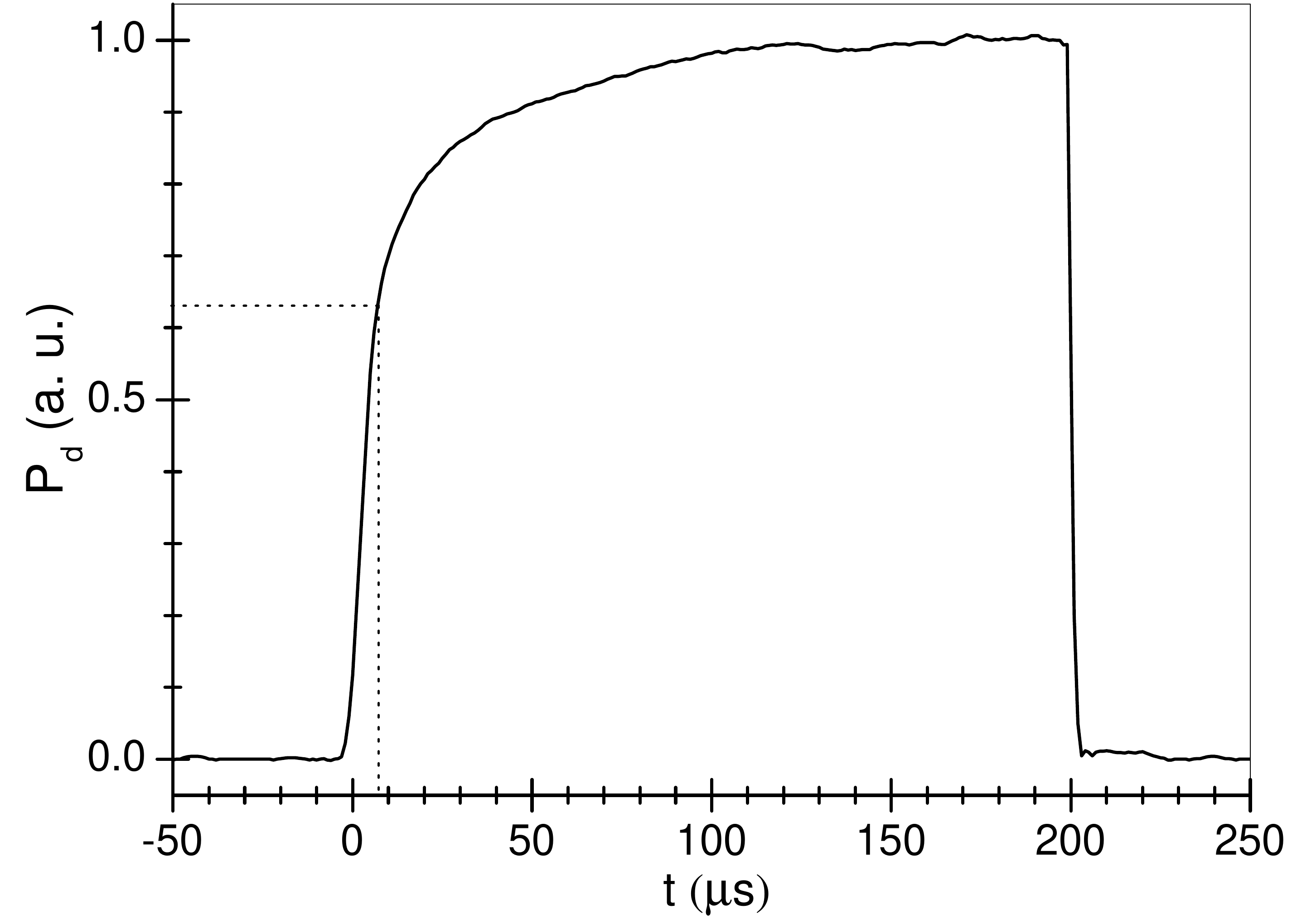}} 
\caption{Measurement of the instability's response time. We measure the temporal evolution of the light power $P_d$ in the orthogonal channel (see text) when the pumps are applied abruptly. The rise time is approximately 7 $\mu$s as indicated by the dotted lines.}
\label{fig3}
\end{center}
\end{figure}

 We have found that with our cold atomic cloud of large optical density (OD), the spin instability does not require the retro-reflected configuration of ref.~\cite{Firth1990} to be observed. In the rest of the paper, we will thus use the ''independent pumps'' setup depicted in Fig.~\ref{fig1}A. The essential features of the instability remain identical in both configurations, the main difference being the ability to tune the spatial length scale of the patterns in the retro-reflected situation~\cite{Firth1990, Ciaramella1993, Firth2016}. However, the independent pumps configuration presents the advantage to be more symmetric (both pumps can be exactly balanced in intensity), and facilitates the detection of the signal (no attenuation by transmission through a mirror as in~\cite{Labeyrie2014, Labeyrie2015}). As emphasized in ref.~\cite{Firth2016}, the price to pay is a higher instability threshold in the independent pumps configuration. In the setup of Fig.~\ref{fig1}A, the two counter-propagating beams are identical in waist ($w = 1.4$ mm), intensity , detuning ($\delta = - 7.8 \Gamma$, where $\Gamma$ is the atomic linewidth) and polarization (linear). The residual magnetic field can be adjusted using three pairs of compensation coils (not shown in Fig.~\ref{fig1}A). As outlined before, the nonlinear interaction generates new light beams with a polarization orthogonal to that of the pumps. We thus detect our patterns in this polarization channel, with the help of a polarizing beam splitter (PBS). As in our previous experiments, the atomic cloud is produced in a large magneto-optical trap (MOT) containing roughly $10^{11}~^{87}$Rb atoms, with a typical diameter of 1 cm~\cite{VLMOT}. We load the MOT for 215 ms, then switch off the MOT's lasers and magnetic field before the image acquisition sequence. However, because of the presence of eddy currents, we typically have to wait for 10 ms for the residual B-field to decay to a value small enough for the compensation coils to null it. During this duration without trapping, the cold clouds expands: as a result, the cloud's OD is reduced from an initial value of 150-200 to a final one of roughly 80. Then, the pump pulse is applied and a pattern image such as shown in Fig.~\ref{fig1}B is recorded. As can be seen, for these parameters (pump intensity $I = 128$ mW/cm$^2$ and detuning $\delta = - 7.8 \Gamma$ resulting in a saturation parameter $s = 0.15$ per beam) the patterns consist of bright stripes on a dark background. Topological defects are observed when stripes end, as emphasized by the circles. Counting these defects will allow us to test the KZM hypothesis.
 
 Before turning to the analysis of KZM, we illustrate in Fig.~\ref{fig2} one of the main features of the spin instability: its sensitivity to the magnetic field. We record the diffracted power $P_d$ detected in the orthogonal channel, as we vary the component $B_{//}$ of the magnetic field parallel to the polarization of the pumps (note the double logarithmic scale). The two others components of the magnetic field are set to zero. The diffracted power corresponds to the light generated in the spatial mode(s) of the instability, and thus does not include the power remaining in the mode of the pumps (the separation between the two is readily achieved using a far-field imaging setup). In Fig.~\ref{fig2}, $P_d$ is normalized to the pump power $P_{in}$ (note however that the absolute vertical scale is not significant). Each curve corresponds to a different $P_{in}$ and thus to a different value of $s$, as indicated in the caption. We checked that these curves are identical when the sign of $B_{//}$ is reversed. We observe the following generic behavior: $P_d$ is maximum around  $B_{//} = 0$, decreases rapidly when $B_{//}$ is increased until the instability vanishes, and then increases again when $B_{//}$ is further increased. This actually corresponds to two different instabilities (''type 1'' and ''type 2'' as indicated in the figure), separated by an intermediate stable region. These instabilities generate strikingly different patterns, and exhibit different thresholds as a function of the various parameters (e.g. pump intensity or cloud's OD). For instance, on curve 1 of Fig.~\ref{fig2} the type 1 instability around $B_{//} = 0$ is absent (because the pump intensity is below the corresponding threshold) while the type 2 instability at ''large'' $B_{//}$ is present. We observe different behaviors when the two other components of the magnetic field are varied. It is worth mentioning here that spatial and polarization instabilities were observed with two counter-propagating beams in early experiments with hot vapors~\cite{Gauthier1990, Grynberg1992}. More recently, Gauthier and co-workers have also reported spatial optical instabilities using a cold atomic cloud in a similar configuration~\cite{Greenberg2012}, and attributed their observations to another mechanism based on Sisyphus cooling-assisted spatial bunching. The comparison between these various instabilities remains to be studied.
 
 The analysis of the rich behavior observed versus magnetic field and the understanding of the various spin instabilities is currently a work in progress, and will be the topic of a forthcoming publication~\cite{Kresic2016}. In the rest of the paper, we discuss the validity of the KZM scenario using the type 1 instability and thus setting the residual B-field to zero. To achieve the sweep of a control parameter across the transition, we linearly ramp the laser intensity during the pump pulse from 0 to $I = 128$ mW/cm$^2$ within a sweep time $t_s$ using an acousto-optic modulator. The pump intensity is then held at $I$ for 10 $\mu$s and an image of the spontaneously generated pattern is recorded. In our experiment, the interval spanned for the sweep time is limited by different factors. As pointed out in~\cite{Arecchi1999}, $t_s$ must be larger than the response time of the system (the ''inertial time''). In our situation, this is typically a Zeeman pumping time which is relatively short (of the order of a few $\mu$s) for our experimental conditions. Indeed, the total saturation parameter associated with the two beams (neglecting absorption) and the employed detuning of $\delta = -7.8~\Gamma$ is of the order of 0.3, corresponding to a photon scattering time of 0.23 $\mu$s. Achieving a significant Zeeman pumping in the stretched states may require the scattering of several tens of photons. Our minimal sweep time of 5 $\mu$s (limited by the pulse generator) roughly corresponds to 20 scattered photons per atom. To estimate more accurately the actual response time of the nonlinear system, we inserted a photodiode in place of the CCD shown in Fig.~\ref{fig1}A, and measured the temporal evolution of $P_d$ when the pumps are applied abruptly. This is shown in Fig.~\ref{fig3}. We typically observe large fluctuations of the curve $P_d (t)$ for successive shots in the experiment. Thus, the curve plotted in Fig.~\ref{fig3} is obtained by averaging $P_d (t)$ over several tens of shots. The time constant extracted from this measurement (corresponding to $P_d = 1-1/e$) is roughly 7 $\mu$s, just above our minimal sweep time. On the long sweeps side, we observed that increasing $t_s$ beyond 150 $\mu$s results in a displacement of the position of the patterns in the beam's transverse profile. This spurious effect could be due to a mechanical effect of the beams on the atoms, or to a non negligible variation of the residual magnetic field when the sweep time is too large. We thus restricted ourselves to sweep times between 5 and 150 $\mu$s.
 
\begin{figure}
\begin{center}
\resizebox{1.0\columnwidth}{!}{\includegraphics{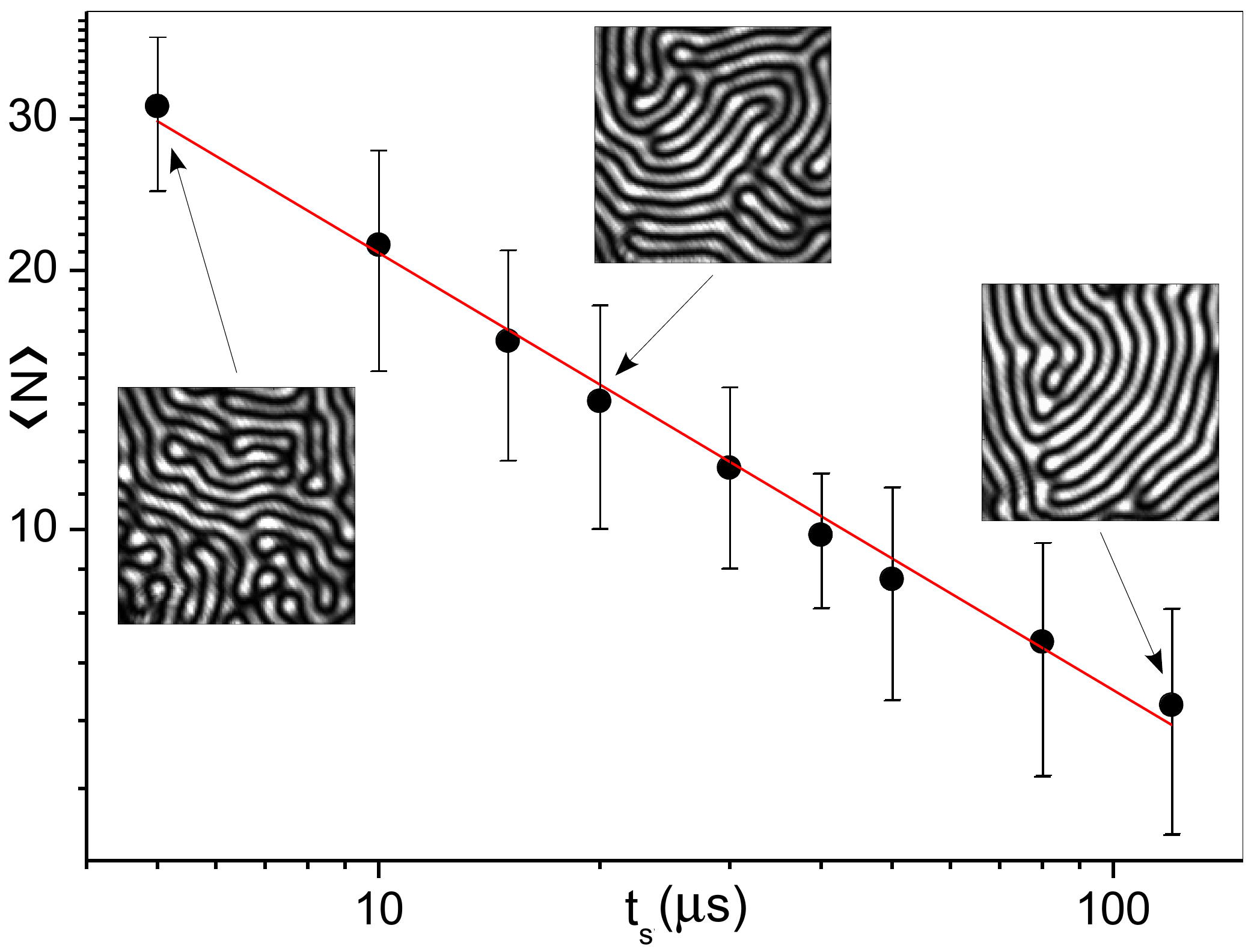}}
\caption{Scaling of the average defect number with sweep time. A power-law decrease $\left\langle N\right\rangle \propto t_s^{-0.51}$ is observed (the line shows the fit). The inserts correspond to typical pattern images for short, intermediate and long sweeps. The peak intensity is $I = 128$ mW/cm$^2$ and the detuning $\delta = -7.8~\Gamma$.}
\label{fig4}
\end{center}
\end{figure}

 The main result of this Letter is shown in Fig.~\ref{fig4}, where we plot the average number of defects $\left\langle N\right\rangle$ versus $t_s$. $\left\langle N\right\rangle$ is obtained using a statistical sample of 20 pattern images such as shown in Fig.~\ref{fig1}B. We observe a clear power-law decrease of the average defect number with increasing sweep duration $\left\langle N\right\rangle \propto t_s^{-\alpha}$. A power-law fit (line) yields an exponent $\alpha = 0.51 \pm 0.015$, riminiscent of the 1/2 value predicted for an overdamped second order phase transition~\cite{Zurek1998}. Indeed, the argument of KZM, initially devised for phase transitions, can be adapted to out-of-equilibrium bifurcations~\cite{Casado2007, Miranda2013}. The prediction $\alpha = 1/2$ also applies to a supercritical bifurcation, the counterpart of an overdamped second order phase transition, and was indeed observed~\cite{Arecchi1999}. For subcritical bifurcations, as e.g. in convective systems, other exponents with smaller absolute values were also found~\cite{Casado2001, Miranda2013}.

In conclusion, we reported in this paper the observation of a spatial instability occurring when two red-detuned counter-propagating laser beams interact inside a large cloud of cold atoms. This instability relies on the spatial modulation of the spin degrees of freedom in the transverse plane, and results in a magnetic self-organization of the atomic sample. We presented an experimental evidence of the Kibble-Zurek mechanism in the formation of topological defects when crossing the threshold of this instability. The measured exponent of 1/2 for the scaling of the defect number versus transition time is consistent with the KZM prediction for a supercritical bifurcation. Other bifurcations, possibly subcritical, could be investigated in the same system, using the same nonlinear mechanism or another for instance the optomechanical mechanism~\cite{Labeyrie2014}. The study of topological defects appears as a convenient tool to investigate critical behaviors, as emphasized e.g. by recent observations in Bose-Einstein condensates where exponents consistent with beyond mean-field theory were reported~\cite{Dalibard2015, Navon2015}.

\acknowledgments{The experiment was performed with the financial support of CNRS, UNS, and R\'{e}gion PACA. We ackowledge many stimulating discussions about spin instabilities with T. Ackemann and colleagues at Strathclyde University (UK).}


\begin{thebibliography}{10}

\bibitem{Kibble1976} T. W. Kibble, J. Phys. A Math. Gen. \textbf{9}, 1387 (1976).

\bibitem{Zurek1985} W. Zurek, Nature (London) \textbf{317}, 505 (1985).

\bibitem{Zurek2002} J. Dziarmaga, A. Smerzi, W. H. Zurek, A. R. Bishop, Phys. Rev. Lett. \textbf{88}, 167001 (2002).

\bibitem{Zurek2000} W. H. Zurek, L. M. A. Bettencourt, J. Dziarmaga, N. D. Antunes, Acta Phys. Polon. \textbf{B31}, 2937 (2000).

\bibitem{Casado2007} S. Casado, W. Gonz\'{a}lez-Vi\~{n}as, S. Boccaletti, P. L. Ramazza, and H. Mancini, Eur. Phys. J. Special Topics \textbf{146}, 87 (2007).

\bibitem{Miranda2013} M. A. Miranda, J. Burguete, H. Mancini, and W. Gonz\'{a}lez-Vi\~{n}as, Phys. Rev. E \textbf{87}, 032902 (2013); M. A. Miranda, D. Laroze and W. Gonz\'{a}lez-Vi\~{n}as, J. Phys.: Condens. Matter \textbf{25}, 404208 (2013).

\bibitem{Yurke1991} I. Chuang, R. Durrer, N. Turok, and B. Yurke, Science \textbf{251}, 1336 (1991).

\bibitem{Arecchi1999} S. Ducci, P.L. Ramazza, W. Gonz\'{a}les-Vi\~{n}as, and F.T. Arecchi, Phys. Rev. Lett. \textbf{83}, 5210 (1999).

\bibitem{Pickett1999} C. B{\"a}uerle, Y. M. Bunkov, S. Fisher, H. Godfrin, and G. Pickett, Nature \textbf{382}, 332-334 (1996).

\bibitem{Monaco2009} R. Monaco, J. Mygind, R. Rivers, and V. Koshelets, Phys. Rev. B \textbf{80}, 180501 (2009).

\bibitem{Singer2013} S. Ulm, J. Rosnagel, G. Jacob, C. Deg{\"u}nther, S. Dawkins, U. Poschinger, R. Nigmatullin, A. Retzker, M. Plenio, F. Schmidt-Kaler and K. Singer, Nature Commum. \textbf{4}, 2290 (2013).

\bibitem{Sadler2006} L. E. Sadler, J. M. Higbie, S. R. Leslie, M. Vengalattore,  D. M. Stamper-Kurn, Nature \textbf{443}, 312-315 (2006).

\bibitem{Weiler2008} C. N. Weiler \textit{et al.}, Nature \textbf{455}, 948 (2008).

\bibitem{Chen2011} D. Chen, , M. White, C. Borries, B. DeMarco, Phys. Rev. Lett. \textbf{106}, 235304 (2011).

\bibitem{Ferrari2013} G. Lamporesi, S. Donadello, S. Serafini, F. Dalfovo, G. Ferrari, Nature Physics \textbf{9}, 656 (2013).

\bibitem{Braun2014} S. Braun \textit{et al.}, PNAS \textbf{112}, 3641 (2015); http://arxiv.org/abs/1403.7199 (2014).

\bibitem{Corman2014} L. Corman \textit{et al.}, Phys. Rev. Lett. \textbf{113}, 135302 (2014).

\bibitem{Navon2015} N. Navon, A. L. Gaunt, R. P. Smith, Z. Hadzibabic, Science \textbf{347}, 167 (2015).

\bibitem{Dalibard2015} L. Chomaz, L. Corman, T. Bienaim\'{e}, R. Desbuquois, C. Weitenberg, S. Nascimb\`{e}ne, J. Beugnon, J. Dalibard, Nature Commum. \textbf{6}, 6162 (2015).

\bibitem{Labeyrie2014} G. Labeyrie, E. Tesio, P.M. Gomes, G.-L. Oppo, W.J. Firth, G.R.M. Robb, A.S. Arnold, R. Kaiser and T. Ackemann, Nature Photonics \textbf{8}, 321 (2014).

\bibitem{Labeyrie2015} A. Camara, R. Kaiser, G. Labeyrie, W.J. Firth, G.-L. Oppo, G.R.M. Robb, A.S. Arnold, and T. Ackemann, Phys. Rev. A \textbf{92}, 013820 (2015).

\bibitem{Firth2016} W.J. Firth, I. Kresic, G. Labeyrie, A. Camara, P.M. Gomes, T. Ackemann, arXiv:1606.04885 [physics.atom-ph].

\bibitem{Grynberg1994} G. Grynberg, A. Ma\^itre and A. Petrossian, Phys. Rev. Lett. \textbf{72}, 2379 (1994).

\bibitem{Ackemann1994} T. Ackemann and W. Lange, Phys. Rev. A \textbf{50}, R4468 (1994).

\bibitem{Ackemann1995} T. Ackemann, Y. Logvin, A. Heuer, and W. Lange, Phys. Rev. Lett. \textbf{75}, 3450 (1995).

\bibitem{Firth1990} W.J. Firth, J. Mod. Opt. \textbf{37}, 151 (1990).

\bibitem{Grynberg1988a} G. Grynberg, Opt. Commun. \textbf{66}, 321 (1988).

\bibitem{Kresic2016} I. Kresic et \textit{al.}, \textit{in preparation}.

\bibitem{Ciaramella1993} E. Ciaramella, M. Tamburrini and E. Santamato, Appl. Phys. Lett. \textbf{63}, 1604 (1993).

\bibitem{VLMOT} A. Camara, R. Kaiser and G. Labeyrie, Phys. Rev. A \textbf{99}, 063404 (2014).

\bibitem{Gauthier1990} D. J. Gauthier, M. S. Malcuit, A. L. Gaeta, and R. W. Boyd, Phys. Rev. Lett. \textbf{64}, 1721 (1990).

\bibitem{Grynberg1992} A. Petrossian, M. Pinard, A. Ma\^{i}tre, J. Y. Courtois and G. Grynberg, Europhys. Lett. \textbf{18}, 689 (1992).

\bibitem{Greenberg2012} J. A. Greenberg and D. J. Gauthier, Phys. Rev. A \textbf{86}, 013823 (2012); J. A. Greenberg and D. J. Gauthier, EPL \textbf{98}, 24001 (2012).

\bibitem{Zurek1998} A. Yates and W. H. Zurek, Phys. Rev. Lett. \textbf{80}, 5477 (1998).

\bibitem{Casado2001} S. Casado, W. Gonz\'{a}lez-Vi\~{n}as, H. Mancini, and S. Boccaletti, Phys. Rev. E \textbf{63}, 057301 (2001).


\end{thebibliography}
\end{document}